\documentclass[apl,twocolumn,showpacs,superscriptaddress]{revtex4-1}

\usepackage{graphicx}
\usepackage{natbib}

\begin{document}

\title{Octahedral tilt independent magnetism in confined GdTiO$_3$ films}

\author{R. F. Need}
 \affiliation{Materials Department, University of California, Santa Barbara, California 93106, USA}
 
\author{B. J. Isaac}
 \affiliation{Materials Department, University of California, Santa Barbara, California 93106, USA}

\author{B. J. Kirby}
 \affiliation{NIST Center for Neutron Research, National Institute of Standards and Technology, Gaithersburg, Maryland 20899 USA}

\author{J. A. Borchers}
 \affiliation{NIST Center for Neutron Research, National Institute of Standards and Technology, Gaithersburg, Maryland 20899 USA}

\author{S. Stemmer}
 \affiliation{Materials Department, University of California, Santa Barbara, California 93106, USA}

\author{S. D. Wilson}
 \email{stephendwilson@engineering.ucsb.edu}
 \affiliation{Materials Department, University of California, Santa Barbara, California 93106, USA}
 
\begin{abstract}
Polarized neutron reflectometry measurements are presented exploring the evolution of ferrimagnetism in GdTiO$_3$ films as they are confined between SrTiO$_3$ layers of variable thicknesses.  As GdTiO$_3$ films approach the thin layer limit and are confined within a substantially thicker SrTiO$_3$ matrix, the TiO$_6$ octahedral tilts endemic to GdTiO$_3$ coherently relax toward the undistorted, cubic phase of SrTiO$_3$.  Our measurements reveal that the ferrimagnetic state within the GdTiO$_3$ layers survives as the TiO$_6$ octahedral tilts in the GdTiO$_3$ layers are suppressed.  Furthermore, our data suggest that a magnetic dead layer develops within the GdTiO$_3$ layer at each GdTiO$_3$/ SrTiO$_3$ interface. The ferrimagnetic moment inherent to the core GdTiO$_3$ layers is negligibly (in models with dead layers) or only weakly (in models without dead layers) impacted as the octahedral tilt angles are suppressed by more than 50$\%$ and the $t_{2g}$ bandwidth is dramatically renormalized.      
\end{abstract}

\maketitle

Complex oxide thin films and interfaces continue to constitute an exciting frontier in condensed matter physics where layer thickness, interfacial strain, and chemistry can be used to tune competing interactions and generate emergent ground states \cite{x,xi}. This tunability, when combined with strong electron-electron correlations in these systems, results in a range of electronic and magnetic ground states unique from their bulk components, such as interfacial ferromagnetism \cite{xii,xiii}, metal-to-insulator transitions \cite{xiv}, and voltage-tunable superconductivity \cite{xv,xvi}.

Within the realm of engineered heterostructures, ABO$_3$ perovskites have received considerable attention, due in part to the wide range of possible chemistries and the atomic precision with which multilayer films can be fabricated.  For many bulk perovskites, the A-site cation is too small for the perovskite structure to retain cubic ($Pm\overline{3}m$) symmetry. The consequence is a cooperative distortion (i.e. tilts and rotations) of the BO$_6$ octahedra network that may take one of multiple possible patterns \cite{xvii} and is proportional in magnitude to the Goldschmidt tolerance factor \cite{xviii}. As the radius of the A-site cation decreases, the structural distortions increase leading to movement of the B-O-B bond angle away from 180$^{\circ}$ and a corresponding decrease in orbital overlap that can effect both electronic and magnetic properties \cite{xxi, xxii, xxiii}. 

These cooperative distortions are altered from their bulk patterns near a heterointerface of two dissimilar perovskite films (i.e. ABO$_3$/A$^{\prime}$B$^{\prime}$O$_3$) \cite{xix}. Which octahedral network distorts and the degree of distortion can be intentionally engineered by the choice layer thicknesses and interfacial strain to generate emergent properties \cite{xx, xxvi}. For example, interfacial octahedral engineering has been successfully employed to enhance ferroelectric polarization in CaTiO$_3$/BiFeO$_3$ superlattices \cite{xxvii}, magnetism in LaMnO$_3$/SrTiO$_3$ superlattices \cite{xxviii}, and to manipulate quantum criticality in SmTiO$_3$/SrTiO$_3$ and GdTiO$_3$/SrTiO$_3$ quantum wells \cite{xliii,xliv}.

Particularly fascinating phenomena appear at engineered GdTiO$_3$/SrTiO$_3$ interfaces.  In the bulk, the Mott insulator GdTiO$_3$ (GTO) possesses GdFeO$_3$-type distortions in its TiO$_6$ octahedra network while the band insulator SrTiO$_3$ (STO) possesses the undistorted parent cubic structure at room temperature \cite{viii, xxiv}. By interfacing thin epitaxial layers of GTO and STO, the octahedral tilts inherent to each layer can be coherently controlled with dramatic effects on the free carriers generated by the polar discontinuity at the interface. For instance, transport measurements have shown that this system goes through a Mott-Hubbard-like metal-to-insulator transition when carriers within STO quantum wells 2 uc (unit cells) thick or less are sandwiched between relatively thick GTO layers \cite{xiv}. In samples with thin GTO layers, SQUID magnetometry has suggested a critical GTO thickness of 6 uc (2 nm), below which GTO transitions from its bulk ferrimagnetic state \cite{vii, xlvi} into a paramagnetic state in conjunction with a 33$\%$ reduction in Ti-O-Ti bond angles in the center of the GTO layers \cite{i}.  This implies an ability to exert fine control over the magnetic state of GTO through interfacial manipulation of its octahedral tilts.

\begin{figure*}[th]
\centering \includegraphics[width =16cm]{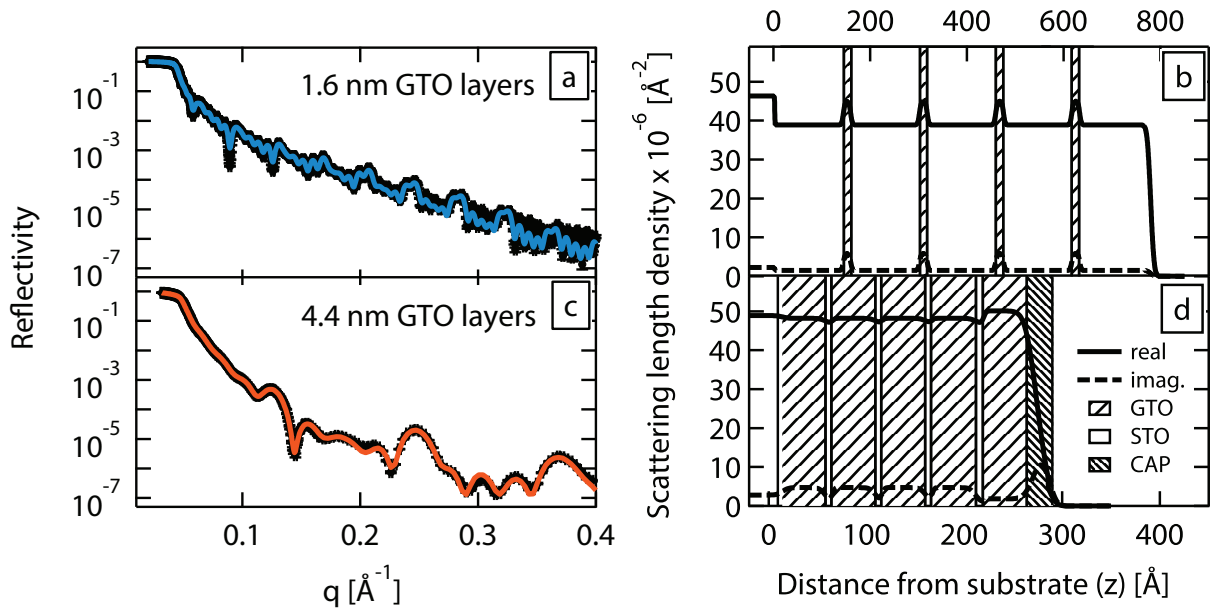}
\caption{X-ray reflectometry data and calculated fits to two representative GTO-STO superlattices measured in this study: (a) 1.6 nm or 4 uc thick GdTiO$_3$ layers, and (c) 4.4 nm or 11 uc thick GdTiO$_3$ layers. The refined models from which the curve fits were calculated are shown in panels (b) and (d) for thin and thick GTO samples respectively.} 
\end{figure*}

In this paper, we explore the coupling between octahedral tilting and magnetism in confined GTO films by using polarized neutron reflectometry (PNR) to probe their interplay in thin GTO layers. Surprisingly, our data show no evidence of a ferrimagnetic-paramagnetic transition near the thin well limit, but rather that GTO remains ferrimagnetic down to layers as thin as 4 uc (1.6 nm). The magnetization curves extracted from the PNR data are analyzed using models both with and without magnetic dead layers (MDLs) in the GTO. When examined using a model with no MDLs, the thinnest GTO layers show $\leq$23$\%$ suppression in the apparent, saturated magnetization. Inclusion of MDLs into the PNR model results in better fits to the experimental data, and a magnetization response that is  independent of GTO layer thickness.  Our results indicate that the substantial relaxation of TiO$_6$ octahedral tilts in GTO/STO interfaces at the thin GTO layer limit has minimal impact on the magnetically ordered state.  More broadly, this implies that ferrimagnetism in GTO is largely independent of the interface-engineered t$_{2g}$ bandwidth.

Superlattice samples of alternating GTO and STO layers were grown for this study using hybrid molecular beam epitaxy as described elsewhere \cite{iv,v,vi}. The degree of distortion/tilting within the GTO titania octahedra network was controlled by varying the thickness of the GTO layers. Previous scanning transmission electron microscopy (STEM) measurements of Gd-O-Gd bond angles are used as a proxy for the relation between layer thickness and octahedra tilts \cite{i}. The thin GTO superlattice contained 4 uc (1.6 nm) GTO layers, in which all of atomic planes within the GTO were distorted from their bulk tilting pattern by 50$\%$ or more. The thick GTO superlattice had 11 uc (4.4 nm) GTO layers, in which the bulk GTO tilt structure was present throughout the entirety of the layers with the exception of the one unit cell at each interface where tilts are suppressed as the titania network transitions into the neighboring STO. For this sample, thin STO spacers (0.6 nm) were used to reduce distortions to the interfacial GTO tilts.

Polarized neutron reflectometry measurements were performed on the PBR reflectometer at the NIST Center for Neutron Research with an incident wavelength of 4.75 $\AA$. Samples were mounted in a cryostat with the film's surface normal to the scattering wavevector, $q$. PNR measurements were collected in a zero field cooled (ZFC) state by cooling the sample from well above the Curie temperature to 5 K under zero field, then polarizing the sample to $\mu_{0}$H = 3 T applied in the plane of the film and collecting PNR scans as the field was stepped back to zero. The layer thicknesses and interface quality of these samples were characterized using non-resonant, unpolarized x-ray reflectometry (XRR) performed with a Cu K$\alpha$ lab diffractometer. XRR measurements were performed in air at room temperature. All reflectometry data sets were refined to slab layer models using the Refl1D code that implements an optical matrix formalism \cite{ii, iii}.

Figures 1(a) and 1(c) contain the XRR data and model fits for thin and thick GTO superlattices, respectively. The refined structural models corresponding to these samples are shown in the Figs. 1(b) and 1(d). During refinement, all layers were allowed to have an independently refined thickness, but layer chemistry and interface roughness were confined to be uniform for all layers of a given type in order to reduce the number of free parameters. The topmost layer in each sample was allowed to be unique in order to account for surface degradation known to occur in rare earth transition metal oxides \cite{xlvii}. Average GTO layer thicknesses were refined to be 4.48(6) nm for the thick GTO sample and 1.57(6) nm for the thin GTO sample and are in near perfect agreement with the designed structures. The apparent chemical roughnesses, which are effectively averaged over the entire x-ray beam spot ($\approx$10 mm$^2$), span a small range from 2.3 $\AA$ - 4.4 $\AA$ and attest to the excellent quality of these films. Previous reflectometry and electron microscopy studies on this system suggest local interfaces are in fact atomically sharp \cite{xiii,xxiv}, and the apparent roughness values arise from steps on the substrate surface propagating upwards through the film, rather than chemical intermixing.

\begin{figure}
\includegraphics[width=8.5cm]{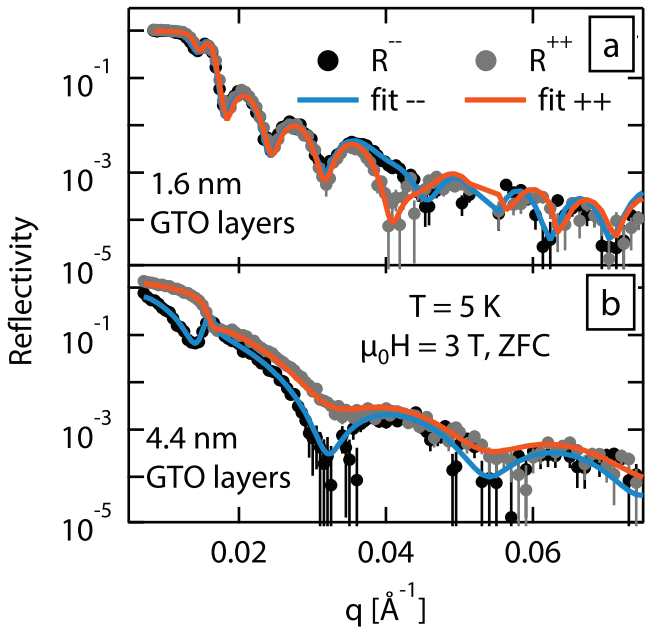}
\caption{Polarized neutron reflectometry data and refined fits for (a) thin GTO layer and (b) thick GTO layer superlattices measured in a ZFC state under a $\mu_{0}$H = 3 T applied field.}  
\end{figure}

We begin by analyzing PNR data for the thin GTO sample with 4 uc GTO layers using a magnetization model without dead layers, similar to that previously applied to the GTO/STO system \cite{xiii}. Figure 2(a) shows PNR data collected at 5 K after cooling under zero field then applying $\mu_{0}$H = 3 T at base temperature. Data refinement shows that the thin GTO layers still exhibit a net in-plane magnetization and reach 2.7(1) $\mu_{B}/$fu in the center of the GTO layers under the assumption of no magnetic dead layers. While this is lower than the 3.5(1) $\mu_{B}/$fu observed in the thick GTO superlattice (Fig. 2 (b)), the 23$\%$ magnetization reduction observed here is significantly less than the 85$\%$ reduction observed using a volume-averaged technique \cite{i}. The survival of bulk-like magnetism at this thin well limit where TiO$_6$ octahedral tilts have been suppressed by over 50$\%$ is surprising \cite{i} and deviates from the current picture of completely quenched magnetism at this limit.  This contradiction in the apparent suppression between depth-resolved and volume-averaged probes suggests the presence of magnetic dead layers that create a finite thickness effect.

\begin{figure}
\includegraphics[width=8.5cm]{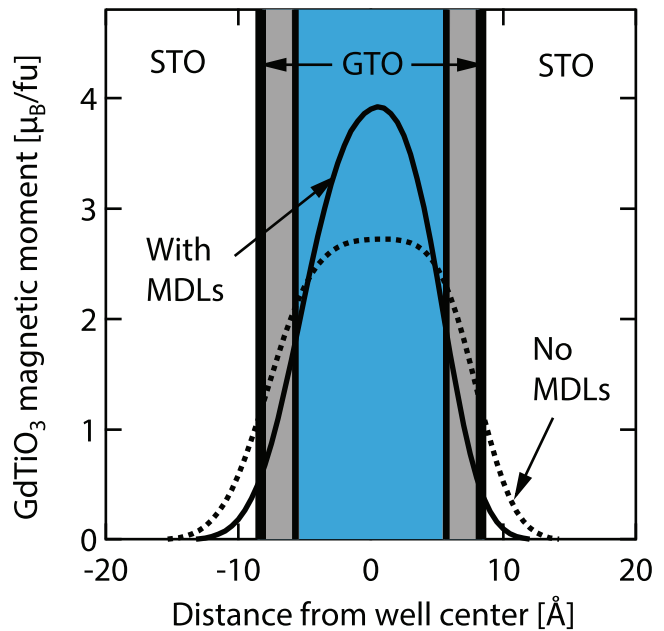}
\caption{Refined GTO layer magnetization profiles for the thin GTO superlattice under the assumptions of no MDLs and 2.5 $\AA$ MDLs. These profiles are overlaid on a schematic representation of the best MDL model, in which the MDLs (grey regions) begin at the chemical GTO/STO interface and extend 2.5 $\AA$ into the GTO layers.}  
\end{figure}

Therefore, the data were reanalyzed incorporating MDLs into the layer model of the multilayer film. A number of different MDL models were compared with the best models providing better visual and numerical fits to the PNR data than models without MDLs \cite{xxxiv}.  The most descriptive model is shown in Fig. 3 where refined magnetization profiles of the thin GTO superlattice with and without MDLs are overlaid on the chemical layers. This model has matching MDLs on both sides of the GTO layer that begin at the chemical GTO/STO interface and extend 2.5 $\AA$ into the GTO layer (i.e. none of the MDL is contained in the STO layers). Roughnesses of the MDLs were constrained to be no smaller than the chemical roughnesses of the interfaces where the MDLs were located. The justification for this roughness constraint stems from the interpretation of the local chemical interface roughnesses arising from the stepped substrate, which implies these values represent lower limits below which roughness values lose physical meaning.  Because of the small MDL thicknesses relative to the chemical roughness, roughnesses were propagated across multiple interfaces when calculating reflectometry profiles. 

Within this model, the moments in the center of the GTO layers increase to 3.9(1) $\mu_{B}/$fu and 3.8(1) $\mu_{B}/$fu for samples with 1.6 nm and 4.4 nm GTO thicknesses, respectively. The larger increase in moment seen in the thin GTO sample highlights the proportional relation between the refined magnetization and the relative volume fraction of GTO layers lost due to the addition of MDLs.

\begin{figure}
\includegraphics[width=8.5cm]{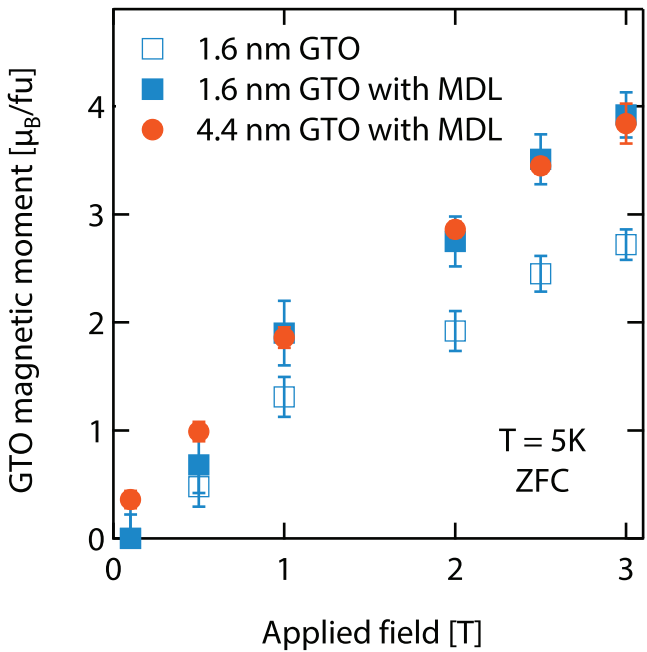}
\caption{GTO magnetic moment values determined via refinement of PNR data measured in a ZFC state. Moments refined with an MDL model are shown by filled symbols. Open symbols show the refined moments for the thin GTO sample when no MDL are included.}  
\end{figure}

Applying the MDL model to the entire ZFC dataset for both thick and thin GTO samples results in a field polarized magnetization that is independent of GTO layer thickness, as shown by the filled symbols in Fig. 4. The thin GTO superlattice refined to a model with no MDLs is also included as a reference. The magnetization data on both films are characterized by little to no remnant magnetization upon field removal and a slow onset of saturation that agree well with previously reported magnetometry data from bulk GTO \cite{vii}. Single ion paramagnetism is ruled out as a possible explanation of this data due to the well-defined order parameter measured in these films \cite{i,xiii} and the temperature dependence of the magnetization that disagreese with predictions from a Brillouin function \cite{xxxiv}.

These combined results suggest that the apparent suppression of magnetization, in this work and also the previous SQUID magnetometry study \cite{i}, is likely an effect of neglecting the magnetic dead layers at the GTO/STO interface and instead averaging magnetization over the entire GTO layer. When these dead layers are incorporated into a model of these systems, the two PNR data sets collapse onto one another, indicating that the magnetism in the center of the GTO layers is independent of the interface-induced octahedral tilting. From the reported bond angles in GTO/STO heterostructures \cite{i}, this is true up to at least a 50$\%$ change in distortion of the octahedral network from its preferred bulk pattern (Ti-O-Ti angle $\approx$ 144$^{\circ}$) towards an undistorted structure (Ti-O-Ti angle = 180$^{\circ}$).  

We stress here that even absent the presence of modeled MDLs, the observed ferrimagnetism in 4 uc thick GTO is only suppressed $23\%$ relative to bulk-like, 11 uc thick GTO.  This is a surprisingly weak perturbation to the magnetism given the known alteration of the octahedral tilt structure in these thin GTO layers and an unambiguous demonstration that robust ferrimagnetism persists well below the previously reported bound of 6 uc thick GTO layers.     

Additional support for the inclusion of MDLs into the model of GTO/STO interfaces comes from the frequency with which heterointerfaces result in the formation of MDLs near the interface. MDLs are often observed in both ferromagnetic metals \cite{xxxi,xxxii,xxxiii} and oxides such as La$_{1-x}$Sr$_x$MnO$_3$ (LSMO) and La$_{1-x}$Ca$_x$MnO$_3$ (LCMO) \cite{xxix, xxx, xxxv, xxxvi, xxxvii}. The origins of these MDLs are typically unique to the interface in question. While structural distortions are a common source of MDLs, that explanation is ruled out in the GTO/STO system because the interfacial, MDL-containing unit cell in thick ($\geq$ 3.5 nm) GTO is distorted by approximately 50$\%$, the same level of distortion that is present in the center of thin (1.6 nm) GTO layers that show unperturbed ferrimagnetism. 

Another possible source of MDLs is orbital reconstruction at the interface. This is particularly relevant for oxide heterostructures where interfacial orbital reconstruction is regularly observed \cite{xl,xli,xlii,xlv}. In the case of thin LSMO layers, x-ray measurements have shown that the 3z$^2$-r$^2$ orbital is preferentially occupied, leading to a weakening of the double exchange responsible for LSMO's FM and resulting in its observed MDLs \cite{xxxviii, xxxix}. In bulk GTO, first principles calculations suggest both orbital ordering and FM are stabilized by a hybridization of the t$_{2g}$-e$_g$ orbitals \cite{xxi}. This hybridization is due to the GdFeO$_3$-type octahedral distortion and, as that distortion is decreased, FM exchange is weakened. Thus while evidence for orbital reconstruction in GTO/STO has yet to be reported, it is possible to speculate towards a case where, either via compressive strain or symmetry breaking at the interface, an orbital reconstruction occurs. This may result in decreased $t_{2g}$-$e_g$ overlap and hybridization pushing the system towards a FM-AFM instability, but this is not directly reflected in the reported Gd-O-Gd bond angles that have been used as a proxy for octahedral tilting and rotations in this study.

In summary, PNR was used to explore the relationship between the cooperative structural distortion of the TiO$_6$ octahedra network and the ferrimagnetic state in GTO thin films. PNR measurements provide evidence that ferrimagnetism in GTO layers survives as the single layer limit is approached.  Specifically, the saturated moment of the ferrimagnetic state in GTO layers as thin as 4 uc is reduced by only $23\%$ relative to bulk-like layers in models neglecting the potential presence of MDLs and becomes identical to bulk-like layers once models incorporating MDLs are used.  Incorporating thin MDLs at GTO/STO interfaces improves refined models of PNR data; however analysis of the data within either approach reveals that the magnetization in the interior of GTO layers (excluding MDLs) is largely independent of changes in octahedral tilts and rotations as measured by Ti-O-Ti bond angles.  Our data curiously point toward a picture of correlated magnetism in GTO which is decoupled from the modified octahedral tilts thought to drive the metal-insulator instability in this compound.

\acknowledgments{
The authors thank B.B. Maranville and A. Green for development of the Refl1D code with roughness propagation. S.W., R. N., and S.S. acknowledge support under ARO award number W911NF1410379. R.N. was supported in part by the National Science Foundation Graduate Research Fellowship under Grant No. 1144085. }

\bibliography{BibTeX_FMinDistortedGTO_APL}

\end{document}